\documentstyle{article}

\font\tenrm=cmr10
\font\tenit=cmti10
\font\elevenbf=cmbx10 scaled\magstep 1
\font\elevenrm=cmr10 scaled\magstep 1
 1

\font\ninerm=cmr9

\renewenvironment{thebibliography}[1]
 { \elevenrm
   \begin{list}{\arabic{enumi}.}
    {\usecounter{enumi} \setlength{\parsep}{0pt}
     \setlength{\itemsep}{3pt} \settowidth{\labelwidth}{#1.}
     \sloppy
    }}{\end{list}}

\begin{document}
\begin{center}{{\elevenbf
           ON THE QUANTIZATION OF $SU(3)$ SKYRMIONS
\footnote{\ninerm\baselineskip=11pt
The work supported by Russian Fund for Fundamental Research, grant
95-02-03868a and by Volkswagenstiftung, FRG}
\\}
\vglue 1.0cm
{\tenrm V.B.KOPELIOVICH \\}
{\tenit Institute for Nuclear Research of the Russian Academy of
Sciences,\\ 60th October Anniversary Prospect 7A, Moscow 117312, Russia\\}
\vglue 0.8cm
{\tenrm ABSTRACT}}
\end{center}
\vglue 0.3cm
{\rightskip=3pc
 \leftskip=3pc
 \tenrm\baselineskip=12pt
 \noindent
The quantization condition derived previously for $SU(2)$ solitons
quantized with $SU(3)$ collective coordinates is generalized for
$SU(3)$ skyrmions with strangeness content different from zero.
Quantization of the dipole-type configuration with large strangeness
content found recently is considered as an example.
 \vglue 0.6cm}
 \vglue 0.1cm
\baselineskip=14pt
\elevenrm

 1. The chiral soliton approach \cite{1} allows not only to describe 
the properties of baryons with a rather good accuracy \cite{2},
\cite{3} but also to make some predictions for the spectrum of states
with baryon number $B>1$ \cite{4}-\cite{6}. The quantization of the 
bound states of skyrmions, their zero modes first of all, is a quite
necessary step towards realization of this approach. Different aspects
of this problem have been considered beginning with the papers \cite{2},
\cite{7} and \cite{4}, however, the complete treatment allowing the
consideration of arbitrary $SU(3)$ skyrmions was absent till now.

In the sector with $B=2$ besides the $SO(3)$ hedgehog with the lowest
quantum states being interpreted as $H$-dibaryon \cite{4} the $SU(2)$
torus - bound $B=2$ state - was discovered almost 10 years ago \cite{6}.
The position of the known $B=2$ classical configurations representing
local minima in $SU(3)$ configuration space is shown on Fig.1 in the 
plane with scalar strangeness content $SC$ as $Y$ axis and the difference 
of $U$- and $D$-contents as $X$-axis. Since the sum of all scalar contents
equals to $1$ they are defined uniquely in each point of this plot.
The $SO(3)$ hedgehog $(1)$ has all contents equal to $1/3$. There
are 3 torii in 3 different $SU(2)$ subgroups of $SU(3)$, the $u-d$
symmetric state $(2)$ with $SC=0$ being of special interest. The dipole
type state $(5)$ found recently \cite{9} has the binding energy about
half of that of the torus.

The quantization of zero modes of solitons has been done previously
in few different cases: for $SU(2)$ solitons rotated in $SU(2)$ and
$SU(3)$ configuration spaces of collective coordinates \cite{2},
\cite{7}, \cite{4}, and also for $SO(3)$ solitons \cite{4}. In the first
case the quantization condition known as Guadagnini's one \cite{7}
was established, see also \cite{10}. The quantization of $SU(2)$ torus 
leads to predictions of rich spectrum of strange dibaryons \cite{11}.
However, these kinds of solitons are only particular cases. Other
types of solitons exist, e.g. solitons of dipole type with large
strangeness content \cite{9}, point $(5)$ on the Fig.1. In view of this 
the quantization procedure of arbitrary $SU(3)$ solitons should be
developed. This is a subject of present paper.\\

 2. Let us consider the Wess-Zumino (WZ) term in the action which defines
the quantum numbers of the system in the quantization procedure. As
usually, we introduce the time-dependent collective coordinates for
the quantization of zero modes according to the relation:
$ U(\vec{r},t)=A(t)U_0(\vec{r})A^{\dagger} (t)$. The integration by
parts is possible then in the expression for the WZ-term in the action
\cite{12}:
$$S^{WZ}=\frac{-iN_c}{240\pi^2} \epsilon_{\mu\nu\alpha\beta\gamma}
 \int_{\Omega}Tr\bar{L}_{\mu}\bar{L}_{\nu}\bar{L}_{\alpha}\bar{L}_{\beta}
\bar{L}_{\gamma} d^5x'  \eqno (1)  $$
$\Omega$ being the 5-dimensional region with the 4-dimensional
space-time as its boundary. We obtain then for the WZ-term contribution
to the lagrangian of the system:
$$L^{WZ}=\frac{-iN_c}{48\pi^2} \epsilon_{\alpha\beta\gamma}\int
Tr A^{\dagger}\dot{A} (U_0L_{\alpha}L_{\beta}L_{\gamma}U_0^{\dagger} +
L_{\alpha}L_{\beta}L_{\gamma}) d^3x \eqno (2) $$
where $\bar{L}_{\mu}=U^{\dagger}d_{\mu}U$, $L_{\alpha}=U_0^{\dagger}
d_{\alpha}U_0=iL_{k,\alpha}\lambda_k$, or
$$L^{WZ}=\frac{N_c}{24\pi^2}\int \sum_{k=1}^{k=8} \omega_k WZ_k d^3x= 
\sum_{k=1}^{k=8} \omega_kL^{WZ}_k \eqno (3) $$
with angular velocities of rotation in the configuration space defined
in usual way, $A^{\dagger}\dot{A}=-{i \over 2} \omega_k\lambda_k$.
Summation over repeated indeces is assumed here and further.
Functions $WZ_k$ can be expressed through the Cartan-Maurer currents
$L_{k,i}$:
$$ WZ_i= (R_{ik}(U_0)+\delta_{ik})\tilde{WZ}_k, \eqno (4)  $$
where
$$\tilde{WZ}_1=-(L_1,L_4L_5+L_6L_7)-(L_2L_3L_8)/\sqrt{3}-
2(L_8,L_4L_7-L_5L_6)/\sqrt{3} $$
$$\tilde{WZ}_2=-(L_2,L_4L_5+L_6L_7)-(L_3L_1L_8)/\sqrt{3}-
2(L_8,L_4L_6+L_5L_7)/\sqrt{3} $$
$$\tilde{WZ}_3=-(L_3,L_4L_5+L_6L_7)-(L_1L_2L_8)/\sqrt{3}-
2(L_8,L_4L_5-L_6L_7)/\sqrt{3} $$
$$\tilde{WZ}_4=-(L_4,L_1L_2-L_6L_7)+(L_3,L_2L_6+L_1L_7)-
(L_8,L_1L_7+L_2L_6+L_3L_5)/\sqrt{3} $$
$$\tilde{WZ}_5=-(L_5,L_1L_2-L_6L_7)-(L_3,L_1L_6-L_2L_7)-
(L_8,L_2L_7-L_1L_6-L_3L_4)/\sqrt{3} $$
$$\tilde{WZ}_6=(L_6,L_1L_2+L_4L_5)-(L_3,L_1L_5-L_2L_4)-
(L_8,L_1L_5-L_2L_4-L_3L_7)/\sqrt{3} $$
$$\tilde{WZ}_7=(L_4,L_1L_2+L_4L_5)+(L_3,L_2L_5+L_1L_4)-
(L_8,L_3L_6-L_1L_4-L_2L_5)/\sqrt{3} $$
$$\tilde{WZ}_8= -\sqrt{3}(L_1L_2L_3)+(L_8L_4L_5)+(L_8L_6L_7) \eqno(5) $$
$(\vec{L}_1\vec{L}_2\vec{L}_3)$ denotes the mixed product of vectors
$\vec{L}_1$, $\vec{L}_2$, $\vec{L}_3$. The real orthogonal matrix
$R_{ik}(U_0)={1 \over 2}Tr\lambda_iU_0\lambda_kU^{\dagger}_0$.

It should be noted that the results of calculation according to $(5)$
depend on the orientation of the soliton in the $SU(3)$
configuration space.

When solitons are located in the $(u,d)$ $SU(2)$ subgroup of $SU(3)$
only $L_1$, $L_2$ and $L_3$ are different from zero, $WZ$ and $\tilde{WZ}$
are both proportional to the $B$-number density and the well known
quantization condition by Guadagnini \cite{7} rederived in \cite{10}
takes place,
$$ Y_R={2 \over \sqrt{3}}dL^{WZ}/d\omega_8 = N_c B/3      \eqno(6) $$
where $Y_R$ is the so called right hypercharge characterizing the
$SU(3)$ irrep under consideration.
This relation will be generalized here to
$$ Y^{min}_R = {2 \over \sqrt{3}} dL^{WZ}/d\omega_8 = 
N_c B (1 - 3 SC)/3      \eqno(7)  $$
This formula was checked for several cases.

a) We can rotate any $SU(2)$ soliton by arbitrary constant $SU(3)$
matrix containing $U_4=exp(-i\nu\lambda_4)$. In this case $SC=
{1 \over2} sin^2\nu$, both $WZ_8$, $\tilde{WZ}_8$ are proportional
to $R_{88}=1 - {3 \over 2}sin^2\nu$. As a result, the relation $(7)$
is fulfilled exactly. Solitons $(3)$ and $(4)$ on Fig.1 can be obtained
from $(u,d)$ soliton $(2)$ by means of $U_4$ or $U_2U_4$ rotations and
satisfy relation $(7)$.

b) For the $SO(3)$ hedgehog $SC=1/3$, \cite{6} and $L^{WZ}=0$, \cite{4}
which satisfies $(7)$ again.

c) We obtained the relation $(7)$ numerically for the solitons of the type

$        U = U_L(u,s) U(u,d) U_R(d,s)  $ \cite{9} with
$        U(u,d)= \exp(ia \lambda_2 ) \exp(ib \lambda_3) $
and $U_L(u,s)$ and $U_R(d,s)$ being deformed interacting $B=1$
$SU(2)$ hedgehogs. For this ansatz we had for rotated $SU(3)$ Cartan-
Maurer currents \cite{9}:
$$ \begin{array}{ll}
\tilde{L}_{1i}=s_ac_al_{3i}, &
\tilde{L}_{2i}=d_ia,  \nonumber \\[10pt]
\tilde{L}_{3i}=(c_{2a}l_{3i} - r_{3i})/2 + d_ib, & \tilde{L}_{4i} 
= l_{1i}c_a, \nonumber \\[10pt]
\tilde{L}_{5i}=c_a l_{2i}, & \tilde{L}_{6i}=l_{1i}s_a+r_{1i}(b), 
\nonumber \\[10pt]
\tilde{L}_{7i}=s_a l_{2i}+r_{2i}(b), & \tilde{L}_{8i}= 
\sqrt{3} (l_{3i}+r_{3i})/2.
\end{array} \eqno (8) $$

$\tilde{L}=TLT^{\dagger}$,
$U_0 = VT$, $V=U(u,s)exp(ia\lambda_2)$, $T=exp(ib\lambda_3)U(d,s)$.
$s_a=sin a$, $c_a=cos a$, etc.,
in terms of $SU(2)$ C-M currents $l_{k,i}$ and $r_{k,i}$ $(i,k=1,2,3)$
and functions $a$ and $b$. In this case only the integral over the
function $\tilde{WZ}_8$ is different from zero $(N_c=3)$:
$$ \frac{1}{2\sqrt{3}\pi^2} \int\tilde{WZ}_8d^3x =\frac{1}{4\pi^2}
\int[(\vec{l}_1\vec{l}_2\vec{l}_3)+(\vec{r}_1\vec{r}_2\vec{r}_3)]d^3x
 = - (B_L+B_R)/2   \eqno  (9)  $$
where $B_L$ and $B_R$ are the baryon numbers located in left
$(u,s)$ and right $(d,s)$ $SU(2)$ subgroups of $SU(3)$. We should
calculate $(3),(7)$ with $WZ_8=(R_{8k}(V)+R_{k8}(T))\tilde{WZ}_k$. The
contribution $-(B_L+B_R)/2$ also appears, together with some additional
terms which turned out to be very small numerically. We obtained
$SC=0.49$ and $Y_R^{min}=-0.96$.

It is natural to suggest that $(7)$ holds for any $SU(3)$ skyrmions.\\

 3. The expression for the rotation energy of the system depending
on the angular velocities of rotations in $SU(3)$ collective 
coordinates space can be written in such a form:
$$ L_{rot} = \frac{ {F_\pi}^2 }{16} (\tilde{\omega}_1^2+
\tilde{\omega}_2^2 ...  + \tilde{\omega}_8^2) $$
$$
+ {1 \over 8e^2} \Biggl\{
\vec{s}_{12}^2 +
\vec{s}_{23}^2 +
\vec{s}_{31}^2 +
\vec{s}_{45}^2 +
\vec{s}_{67}^2 +
{3 \over 4} \biggl(
\vec{s}_{48}^2 +
\vec{s}_{58}^2 +
\vec{s}_{68}^2 +
\vec{s}_{78}^2 \biggr) + $$
$$ + {1 \over 4} \biggl(
\vec{s}_{46}^2 +
\vec{s}_{47}^2 +
\vec{s}_{56}^2 +
\vec{s}_{57}^2 +
\vec{s}_{14}^2 +
\vec{s}_{15}^2 +
\vec{s}_{16}^2 +
\vec{s}_{17}^2 + 
\vec{s}_{24}^2 +
\vec{s}_{25}^2 +
\vec{s}_{26}^2 +
\vec{s}_{27}^2 +
\vec{s}_{34}^2 +
\vec{s}_{35}^2 +
\vec{s}_{36}^2 +
\vec{s}_{37}^2 \biggr) + $$
$$ + {{\sqrt{3}} \over 2} \biggl(
\vec{s}_{84}
\bigl(
\vec{s}_{16} +
\vec{s}_{34} -
\vec{s}_{27} \bigr) +
\vec{s}_{85}
\bigl(
\vec{s}_{17} +
\vec{s}_{26} +
\vec{s}_{35} \bigr) +
\vec{s}_{86} \bigl(\vec{s}_{14} +\vec{s}_{25} -\vec{s}_{36} \bigr) +
\vec{s}_{87} \bigl(\vec{s}_{15}-\vec{s}_{24}-\vec{s}_{37}\bigr)\biggr)+ $$
$$ + {3 \over 2} \Bigl(
\vec{s}_{12} \bigl(
\vec{s}_{45} +
\vec{s}_{76} \bigr) +
\vec{s}_{23} \bigl(
\vec{s}_{47} +
\vec{s}_{65} \bigr) + 
\vec{s}_{13} \bigl(
\vec{s}_{64} +
\vec{s}_{75} \bigr) +
\vec{s}_{45} \vec{s}_{67} \Bigr) \Biggr\} \eqno (10) $$

Here $\vec{s}_{ik}=\tilde{\omega}_i \vec{L}_k - \tilde{\omega}_k
\vec{L}_i $, $i,k=1,2...8$ are the $SU(3)$ indeces.
The expression for static energy can be obtained from $(10)$ by means of 
substitution $\vec{s}_{ik}=2[\vec{L}_i\vec{L}_k]$, \cite{9}. 
The functions $\tilde{\omega}_i$ are connected with the body fixed
angular velocities of $SU(3)$ rotations by means of transformation
(see $(8)$ above):
$$ \tilde{\omega}=V^{\dagger} \omega V - T \omega T^{\dagger}, $$ 
or
$$ \tilde{\omega}_i=(R_{ik}(V^{\dagger})-R_{ik}(T))\omega_{k}=
R_{ik}\omega_k \eqno(11)$$
$R_{ik}(V^{\dagger})=R_{ki}(V)$ and $R_{ik}(T)$ are real orthogonal
matrices, $i,k=1,...8$. For example,

$ R_{81}=-{\sqrt{3} \over 2}s_{2a}(f_1^2+f_2^2)$, 
$R_{82}=0$, $R_{83}=-{\sqrt{3} \over 2}(c_{2a}(f_1^2+f_2^2)+q_1^3+q_2^2) $

$ R_{85}=-\sqrt{3}c_a(f_0f_1-f_2f_3) $, $R_{88}= {3 \over 2}
(q_1^2+q_2^2-f_1^2-f_2^2) $   , etc                     $(12)$,\\ 

$\bar{U}(u,s)=f_0+i\bar{\tau}_kf_k$, $\tilde{U}(d,s)=q_0+i\tilde{\tau}_kq_k$,
$k=1,2,3$, $\bar{\tau}$ and $\tilde{\tau}$ are the Pauli matrices 
corresponding to $(u,s)$ and $(d,s)$ $SU(2)$ subgroups.

$8$ diagonal moments of inertia and $28$ off-diagonal define the rotation 
energy - quadratic form in $\omega_i\omega_k$ - according to $(10),(11)$.

For the quantization of $SU(2)$ hedgehog in the $SU(3)$ collective 
coordinates space only two different moments of inertia entered \cite{7},
\cite{4}: $\Theta_1=\Theta_2=\Theta_3$ and $\Theta_4=\Theta_5=\Theta_6=
\Theta_7$. For the $SO(3)$ hedgehog the rotation energy also depends
on 2 different inertia: $\Theta_2=\Theta_5=\Theta_7$ and $\Theta_1=
\Theta_3=\Theta_4=\Theta_6=\Theta_8$ \cite{4}.

In the case of strange skyrmion molecule we obtained 4 different
diagonal moments of inertia: $\Theta_1=\Theta_2=\Theta_N$; $\Theta_3$;
$\Theta_4=\Theta_5=\Theta_6=\Theta_7=\Theta_S$ and $\Theta_8$. Numerically
the difference between $\Theta_N$ and $\Theta_3$ is small and both
are about twice smaller than $\Theta_S$. $\Theta_8$ is a bit greater than
$\Theta_S$ (see Table). In view of symmetry properties of the 
configuration many off-diagonal moments of inertia are equal to zero.
Few of them are different from zero, but at least one order of
magnitude smaller than diagonal inertia: $\Theta_{38},\ \Theta_{46},
\ \Theta_{57}$. By this reason we shall neglect them here for the
estimates.
The hamiltonian of the system can be obtained by canonical
quantization procedure \cite{2},\cite{7},\cite{4} (we take the angular
momentum $J=0$) in such simplified form:
$$E_{rot} = \frac{C_2(SU_3) - 3 Y_R^2/4}{2 \Theta_S} +
\frac{N(N+1)}{2} \bigl({1 \over \Theta_N} - {1 \over \Theta_S}
\bigr) + \frac{3 (Y_R - Y^{min}_R)^2}{8 \Theta_8}  \eqno   (13)  $$
$C_2(SU_3)={1 \over 3}(p^2+q^2+pq)+p+q$, $N$ is the right isospin
(see Fig.2), $p,q$ are the numbers of the upper and low indeces
in the tensor describing the $SU(3)$ irrep $(p,q)$.

It is clear from this expression that for $\Theta_8 \rightarrow 0$ 
the right hypercharge $Y_R=Y^{min}_R={2 \over \sqrt{3}}L_8^{WZ}$,
otherwise the quantum correction due to $\omega_8$ will be infinite.
For solitons located in $(u,d)$ $SU(2)$ $\Theta_8=0$ and 
$Y_R={2 \over \sqrt{3}}L_8^{WZ}=B$ - the 
quantization condition \cite{7},\cite{10} with $N_c=3$.

For the skyrmion molecule found in \cite{9} $L^{WZ}_8 \approx
-{\sqrt{3} \over 2}$, or $Y^{min}_R \approx -1$, as it was explained 
above. The last term in $(13)$ is absent for $Y_R=-1$, and because of
the evident constraints
$$\frac{p+2q}{3} \geq Y_R \geq -\frac{q+2p}{3}  \eqno (14) $$
the following lowest $SU(3)$ multiplets 
are possible: octet, $(p,q)=(1,1)$, decuplet $(3,0)$ and antidecuplet
$(0,3)$, Fig.2. 
The sum of the classical mass of
the soliton and rotational energy for the $B=2$ octet, $10$ and $\bar{10}$
is equal to $\sim4.44$, $5.0$ and $5.5$ $Gev$ for $Y_R=-1$. The octets with
$Y_R=0$ and $1$ have masses $4.9$ and $5.0$ $Gev$.
This should be compared with central values of masses of $B=1$
octet and decuplet $2.64$ and $3.05 Gev$ \cite{3}. 
The absolute values of the masses of both $B=1$ and $2$ states are
controlled by the Casimir energies which make contribution of $N_c^0$
into the masses of configurations \cite{13}-\cite{16}. However, the
dipole-type configuration does not differ much from the $B=2$ 
configuration in the product ansatz which we used as a starting one
in our calculations \cite{9}. By this reason the Casimir energy of
the $B=2$ dipole should be close to twice of that for $B=1$ soliton,
and will be canceled in the binding energies of dibaryons. We can
conclude therefore that most of the $B=2$ octet and decuplet states 
should be bound. The nonstrange state appears for the first time 
within the antidecuplet and is unbound.

The mass splittings inside multiplets are defined as usually by flavor
symmetry breaking $(FSB)$ terms in the lagrangian. In this case,
since we start from the soliton with $SC \approx 0.5$, the $FSB$
terms are squeezed by a factor about $\sim 3$ due to the smaller 
dimensions of the kaon cloud in comparison with the pion cloud \cite{9},
and the mass splittings are within $\sim 200-300$ $Mev$. More detailed
calculations will be presented elsewhere.\\

 4. To conclude, the quantization scheme for the $SU(3)$ skyrmions is
presented and the quantization condition found previously \cite{7}
is generalized for skyrmions with arbitrary strangeness content. The
relation $(7)$ is valid for all known $B=2$ local minima in $SU(3)$
configuration space shown in Fig.1. The moments of inertia of arbitrary
$SU(3)$ skyrmions can be calculated with the help of formulas $(10),
(11)$.

For the dipole-type configuration with $SC=0.5$ our results are in 
qualitative agreement with those obtained in \cite{17} for interaction
potential of two strange baryons located at large distances. The new
branch of strange dibaryons additional to known previously \cite{4},
\cite{11} is predicted with smallest uncertainty in the absolute values of
masses due to the Casimir energy, relative to the corresponding $B=1$
states. The prediction by chiral soliton models of the rich spectrum
of baryonic states with different values of strangeness remains one of
the intriguing properties of such models. It is difficult to observe
these states, especially those which are above the threshold for the
decay due to strong interaction. However, further investigations of the
predictions of effective field theories providing new approach of the
description of fundamental properties of matter are of interest.

I am indebted to B.Schwesinger for useful discussions and suggestions
on the initial stages of the work, to B.E.Stern for help in
numerical computations and to G.Holzwarth and H.Walliser for their 
interest in the problems of $SU(3)$ skyrmions.

\vglue 0.4cm
{\elevenbf\noindent References}
\vglue 0.2cm

\newpage \vglue 1in
\vglue 1.0in

\begin{table}

:::::::  $B$  ::: $M$ :: $\Theta_N$:::: $\Theta_S$ ::: $\Theta_3$ ::: 
$\Theta_8$ :: $\Theta_{38}$\\

$FS$ :: 1 : 1702: 3.95 : 1.62 :  -- : : :  -- : :  --  \\
$FS$ :: 2 : 3334: 2.91 : 5.19 : 1.99: 5.71: 0.47 \\

$FSB$: 1 : 1982: 1.98 : 0.73  :  -- : : :  -- : :  --  \\
$FSB$: 2 : 3900: 1.44 : 2.69  : 0.99: 2.85: 0.19 \\
\end{table}
Table. The values of masses (in Mev) and moments of inertia
(in $10^{-3}$ $Mev^{-1}$) for the hedgehog with $B=1$ and dipole
configuration with $B=2$ \cite{9} in flavor symmetric $(FS)$ and
flavor symmetry broken $(FSB)$ cases. $F_{\pi}=186$ $Mev$, $e=4.12$. \\

\vglue 1.0in
{\elevenbf Figure captions} 
\vglue 0.4cm
Fig.1 The position of different classical configurations with
$B=2$ in the plane $(UC-DC)$, $SC$. $UC$, $DC$ and $SC$ are scalar 
quark contents of the soliton.  $UC=(1-Re U_{11})/(3-ReU_{11}-
ReU_{22}-ReU_{33})$, etc. $U_{ii}$ are the diagonal matrix elements
of the unitary matrix $U$. $(1)$ is the $SO(3)$ hedgehog, $(2)$,$(3)$
and $(4)$ are $SU(2)$ torii in $(u,d)$, $(d,s)$ and $(u,s)$ subgroups
of $SU(3)$, $(5)$ is the dipole-type configuration found recently. \\
 
Fig.2 $T_3-Y$ -diagrams for the lowest $SU(3)$ multiplets allowed
for the case of $SU(2)^3$ configurations: octet $(1,1)$, decuplet
$(3,0)$ and antidecuplet $(0,3)$. Dashed line indicates isomultiplets
with $Y=Y^{min}=-1$, $T=N$.


\begin{thebibliography}{17}
\bibitem{1} T.H.R.Skyrme, Proc.Roy.Soc. A260,127(1961) \\
Nucl.Phys. 31,556(1962)
\bibitem{2} G.Adkins, C.Nappi, E.Witten, Nucl.Phys. B228,552(1983) \\
G.Adkins, C.Nappi, Nucl.Phys. B233,109(1984)
\bibitem{3} G.Holzwarth, B.Schwesinger, Rep.Prog.Phys. 49,825(1986) \\
B.Schwesinger, H.Weigel, Phys.Lett. B267,438(1991)
\bibitem{4} A.P.Balachandran et al., Nucl.Phys. B256,525(1985)
\bibitem{5} J.Kunz, P.J.Mulders, Phys.Lett. 215B,449(1988)
\bibitem{6} V.B.Kopeliovich, Yad.Fiz. 47,1495(1988); ibid.51,241(1990);\\
Phys.Lett. 259B,234(1991); Yad.Fiz. 56,260(1993) 
\bibitem{7} E.Guadagnini, Nucl.Phys. B236,35(1984)
\bibitem{8} V.B.Kopeliovich, B.E.Stern, Pis'ma v ZhETF, 45,165(1987)
(JETP Lett. 45,203(1987))\\
J.J.M.Verbaarschot, Phys.Lett. 195B,235(1987)
\bibitem{9} V.B.Kopeliovich, B.E.Schwesinger, B.E.Stern,
Pis'ma v ZhETF, 62,177(1995) (JETP Lett. 62,195(1995)
\bibitem{10} D.I.Dyakonov, V.Yu.Petrov, LNPI Preprint 967(1984)
\bibitem{11} V.B.Kopeliovich, B.E.Schwesinger, B.E.Stern,
Phys.Lett. 242B,145(1990);  Nucl.Phys.A549,485(1992)
\bibitem{12} E.Witten, Nucl.Phys. B223,422,433(1983)
\bibitem{13} I.Zahed, A.Wirzba, U-G.Meissner, Phys.Rev. D33,830(1986)
\bibitem{14} R.V.Konoplich, A.E.Kudryavtsev, R.V.Martemyanov, S.G.Rubin,
Hadron. Journ. 11,271(1988)
\bibitem{15} B.Moussalam, Ann. of Phys. (N.Y.) 225,264(1993)
\bibitem{16} G.Holzwarth, H.Walliser, Nucl.Phys. A587,721(1995)
\bibitem{17} B.Schwesinger, F.G.Scholtz, H.B.Geyer, Phys.Rev. 
D51,1228(1995)
\end{thebibliography}
\end{document}